        \documentstyle[aps,floats]{revtex}
    	\input{psfig}
        
\begin{document}
        \title{Understanding\\ Something About Nothing:\\
        Radiation Zeros}

        \author{Robert W. Brown\thanks{Certain work behind this paper
    has been supported in
        part by the NSF and the CWRU industrial problem solving group.}}
        \address{Physics Department\\Case Western Reserve University\\
        Cleveland Ohio 44106}

        \maketitle

        \begin{abstract}
       Radiation symmetry is briefly reviewed, along with its
    historical, experimental, computational, and theoretical relevance. A
    sketch of the proof of a theorem for radiation zeros is used to
    highlight the connection between gauge-boson couplings and Poincare
    transformations.  It is emphasized that while mostly bad things happen to
good
    zeros, the weak-boson self-couplings continue to be intimately tied to
    the best examples of exact or approximate zeros.
        \end{abstract}

        \section*{Introduction}

        There are radiation zeros all over the place.  See Fig. 1.  Almost all
Born
        amplitudes for the radiation of photons and gluons and other massless
gauge
        bosons have such zeros.  One reason no one notices them,
    however, is that only
        a small fraction occur in physical regions of scattering or decay.
    See Fig. 2 later on.

    \begin{figure} 
        \caption{Even in the early days of QED, zeros could have been
    found.}
        \end{figure}

        The conditions for physical ``null zones'' are sufficiently
        restrictive that the first examples of radiation zeros (the
    Mikaelian-Samuel-Sahdev zeros\cite{MSS}), were discovered only when
        radiative weak-boson production was analyzed.
    (Striking dips in the angular distributions of
    $\overline{\nu}e\rightarrow W\gamma$ and $q\overline{q}\rightarrow W\gamma$
    corresponding to the zeros had been seen earlier.\cite{BSM})
    The zero in the $q\overline{q}$ channel has
    proven to be an important signature in testing the
    trilinear $WW\gamma$ couplings, as wonderful progress
    is drawing tighter and tighter bounds.  So far, all are converging to the
    standard model predictions.\cite{EXP}  The progress has been
    accelerated by theoretical support by Baur,
    Errede, and Landsberg\cite{BEL}, showing that laboratory rapidity
    correlations involving the photon and the charged decay lepton display a
    pronounced dip if the radiation zero exists.  In a very nice talk,
    Tao Han has shown us the
    details (and plots) at this conference.\cite{TH}

    \paragraph*{Additional history.}
    Permit me to digress for a moment.  Our original work\cite{BM,BSM}
    on the production of electroweak pairs in proton collisions
    \begin{equation}
    p \overline{p},\; pp \rightarrow WW,ZZ,WZ,W\gamma
    \end{equation}
    took place in the late $70'$s. This was presented as an alternative to
    electron colliders, betting that the necessary
    energy threshold would first be reached by proton machines.  And
    the theoretical discovery of the first radiation
    zero came in the midst of looking at the corresponding Born calculations.
    So it is gratifying, after all these years,
    to be at the present conference where many experimental
    results, including those pertaining to the $W\gamma$ zero,
    have now been achieved.

    In view of the role of the $W\gamma$ channel,
    I have been given the opportunity to review the theory of the radiation
    zero, and to connect it to another signature of interest, ``approximate''
    zeros. Owing to the fact that out of all the people from all the
    old collaborations from CWRU, Oklahoma
    State, Wisconsin, and SLAC, I happen to be the only one present means I
    have some responsibility to mention the early work.\footnote{I was asked
    by a young CWRU graduate student whether this meant all the others had
passed
    away.}  This is also a welcome chance to talk about a theoretical
    development that, because
    of the lack of examples, is not well known, but perhaps should be.
    Continuing experimental progress provides a good incentive.

    In particular, I would like to emphasize a picture that emerges in the
    proof of general radiation zero theorems.  The emission, or absorption,
    of a gauge boson can be viewed as a local Lorentz transformation (or
    better, a Poincare transformation) of the particle doing the
    emitting or absorbing. This may be of importance in future model
    building, especially since our only successful theories to date,
    gauge theories, lead to universal forms
    for this transformation.

    \section*{Radiation Interference Theorems}

    It seems to me that the experimental, computational and
    theoretical consequences of radiation zeros and their generalizations
    are easiest to describe if we first look at an archetypal theorem for
    their existence and location.  The classical limit of such zeros will
    also be evident.
    The primary example of the general set of radiation zero theorems found
    by Brodsky, Kowalski, and meeski\cite{BB,BKB} is for the emission of
    single photons.

    \subsection*{Single-photon theorem}
    The theorem is the following.  Consider the quantum amplitude
    $M_{\gamma}$ for the emission of a single photon with momentum q.
    Besides
    the photon, assume there are $n$ other
    particle legs ($k$ particles in and $n-k$ particles plus $\gamma$ out,
say).  The
    theorem is that the tree amplitude approximation vanishes, independent
    of anybody's spin, for common charge-to-light-cone-energy ratios, viz.
    \begin{eqnarray}
    M_{\gamma}(tree) = 0\nonumber\\
    {\rm if}\; \frac{Q_i}{p_i \cdot q} = {\rm same,\; all}\; i
    \end{eqnarray}
    where the $i^{th}$ particle has electric charge $Q_i$ and four-momentum
    $p_i$.  The stipulations are that all couplings must be
    ``gauge-theoretic''.  That is, the photon couplings to the particles
    must be as prescribed by local gauge theory, and any derivative
    couplings among the particles themselves must be gauge covariant.
    Also, scalar, spinor, and vector particles can be accommodated
    (spins $\leq 1$).  We return at the end to the question
    of higher spins.

    \subsection*{Physical null zones}

    The factors $\frac{Q_i}{p_i \cdot q}$ come from the coupling
    and particle propagator denominator.  It is easy to go into the complex
    plane to make them equal, and hence zeros are all over the (complex)
    place.  But for them to be equal in
    physical phase space, the first and obvious requirement is that
    all charges must have the same sign (since $p\cdot q \ge 0$).  This
    knocks out many reactions.  By the way, there are only $n-2$
    independent equations.  Interestingly and importantly,
    the last ratio is automatically
    the same if the rest are equal, by virtue of charge and momentum
    conservation.\footnote{What I'm saying is,
    if you have 10\% rotten apples and 10\% rotten
    oranges, then the overall percentage of rotten fruit is the same 10\%.}

    In the garden variety reactions, we could look, for example,
    at electron-electron bremsstrahlung,
    $e^- e^- \rightarrow e^- e^- \gamma$.
    The Born amplitude vanishes if the photon is at right angles to the
    c.m. beam direction, and the final electrons have the same energy.
    The null zone is two-dimensional, and the reason this radiation zero was
    not noticed in radiative corrections calculations is that the
    final-state phase space is sufficiently high dimensional.


    \subsection*{Relevance}

    \paragraph*{EXPERIMENTAL}  The null zone conditions explain why it took
    so long to discover radiation zeros.
    {\it We had to wait for fractionally
    charged quarks and weak bosons
    in order to get three things:  Same-sign fermion-antifermion pairs, a
process
    well-approximated by a Born amplitude, and only three particles plus the
    photon so the
    null zone was simple.}  The amplitude for $q\overline{q} \rightarrow W
    \gamma$ vanishes at a photon scattering angle determined by the quark
charges.  (The $\nu \overline{e} \rightarrow W
    \gamma$ amplitude vanishes at the edge of phase space and is easily
    misinterpreted as a helicity constraint.)  The
    corresponding null zone in radiative $W$ decay\cite{GM} is a line in
    the Dalitz plot.

    The first is still the best example.
    Again, I can refer to Tao Han's talk, but the relevance is that only for
    the very restrictive couplings coming out of gauge theory does the zero
    occur.  In view of the results of many calculations with anomalous
    couplings, and the proof (see below) of the pertinent theorem, I do not
    think it would be hard to put together another theorem.  Accepting the
    present particle content of our world, {\it only in the standard
    electroweak theory will there be a zero in
    the $W\gamma$ reaction.} In any other theory including composite models
    the rapidity dip will get filled in.  We say such experiments test gauge
    theories
    and test the electromagnetic properties of weak bosons.  $\; \: $ :)

    What about other photonic zeros?  The aforementioned zero in
electron-electron
    bremsstrahlung is less interesting as a test, and involves more
    particles.  There has been work on electron-quark and
    quark-antiquark bremsstrahlung, as tests of quark properties, but these
    lead to jet identification experiments, along with the more complicated
    phase space.\cite{BIL,COU,RLS,DH}  We will come back to the subject
    of more tests of weak-boson properties in a bit.
    $\; \: $   :$\vert$

    But as noted most zeros are unphysical. $\; \: $ :(

    \paragraph*{COMPUTATIONAL}  As a by-product of the general proof of
radiation zeros, we learn how to
    rewrite the set of Feynman tree diagrams as a smaller number of factored
    terms, separately gauge-invariant.  It is possible to combine the CALKUL
    photon polarization vectors, for which the fermionic degrees of freedom
    are used as the bases, and very much simplify the amplitude analysis.

    The radiation symmetry may be used, as gauge invariance is used, to
    check the increasingly laborious calculations used in higher-order
    perturbation theory studies.  There is an analogous non-Abelian radiative
    symmetry that exists to check QCD jet calculations, for example.

    \paragraph*{THEORETICAL}  The mechanism in the radiation zero phenomena
that has rather uniquely
    shown itself is the relation between a photon coupling and local Poincare
    invariance.  We repeat that only gauge
    theory couplings lead to universal space-time transformations, and it is
    of value to sketch the proof in order to see how this connection is
    exposed and what universality means.

    \section*{Proof Highlights}

    We can describe the proof fairly succinctly in terms of photon attachments
in a minute.  But first let us lay some groundwork.

    \subsection*{Vertex source graphs}

    Define a ``source'' graph $T_G$ with which we will generate the
    complete tree radiation amplitude by the attachment of a photon in all
possible
    ways.  After simplification of the product of vertex and propagator for the
    attachment of a photon to a given leg, the complete tree
    for a source graph $V_G$ made up of a single vertex (no internal lines) is
    \begin{equation}
    M_{\gamma}(V_G) = \sum \frac{Q_i J_i}{p_i\cdot q}
    \end{equation}
    where we see the $Q/p\cdot q$ factors emerge from the coupling and
    propagator denominator.  The coefficient ``vertex current'' $J_i$ is the
    result of inserting the current $j_i$ into the $i^{th}$ leg, with
    \begin{equation}
    j= j_{conv}+j_{spin}+j_{cont}+j_{YM}
    \end{equation}
    The convection current is $p\cdot \epsilon$,\footnote{Sign changes in
    going from final to initial legs are left understood throughout these
    formulas.}
    the spin current is a first-order momentum-space Lorentz
    transformation of the wave function (e.g., $\frac{i}{4}\sigma^{\alpha
    \beta} \omega_{\alpha \beta}$ for a spinor), the contact current is
    the corresponding transformation of single derivative couplings
    and the Yang-Mills current has the form of $\omega_{\mu \nu} \times$
    (the q terms in the Yang-Mills vertex). The first-order Lorentz parameter
    tensor is
    \begin{equation}
    \omega_{\mu \nu} = q_{\mu}\epsilon_{\nu}-\epsilon_{\mu}q_{\nu}
    \end{equation}
    where $\epsilon$ is the photon polarization vector.

    The key identity is that the sum over the vertex currents is zero
    \begin{equation}
    \sum J_i = 0
    \end{equation}
    because the sum over the convection currents vanishes by momentum
    conservation, the sum over the combination of spin currents and contact
    currents vanishes by Lorentz invariance, and the sum over the Yang-Mills
    current vanishes by the Bianchi identity.
    Equal $Q/p\cdot q$ factors can be pulled out of the sum Eq.(3) and the
    complete radiative process for a vertex source graph vanishes by Eq.(6).

    \subsection*{Allowed couplings}

    The vertex graph could have involved any number of particles, but the
    universal forms of the various currents $j_i$ are preserved only if there
    are important restrictions on any derivative couplings present.  There
    can be no derivatives of Dirac fields; single derivatives of scalar fields
    are allowed; single derivatives of vector fields and double derivatives of
    scalar fields are allowed but only in Yang-Mills trilinear form
    (reminding us of the relationship between the longitudinal vector boson
    and the Goldstone bosons in spontaneous symmetry breaking).
     And the photon couplings also must follow the gauge algorithm:  All
    derivatives are replaced by covariant derivatives.

    \subsection*{Connection to space-time symmetry}

    A succinct way of looking at the previous result is that the attachment
    of a photon generates transformations.  The convection current corresponds
    to a (first-order) displacement of that leg's wave function, the spin
    current to its Lorentz transformation, and the
    contact current to a Lorentz transformation of the derivative coupling.
    For common $Q/p\cdot q$, all these transformations correspond to the
    same universal element and they cancel by invariance.  The photon
attachment
    is a first-order Poincare transformation on the vertex amplitude,
    but by the invariance of that amplitude, it becomes ``unattached''
    in the null zone.

    \subsection*{General source graphs}

    To finish the proof, consider general tree source graphs $T_G$ with (fixed)
    internal lines.  In tacking a photon in all possible ways onto
    the source graph, the new ingredient is internal line attachment.  But
    there are Ward-like identities\footnote{For a scalar line, the identity
    is the same as the Ward identity.} that simplify the problem and are
    of the generic form (D denotes propagators)
    \begin{equation}
    D(p-q)\Gamma D(p) + ({\rm seagulls, if\; any}) = D(p-q)j\frac{Q}{p\cdot q}
+
    \frac{Q}{p\cdot q}jD(p)
    \end{equation}
    Note that we get exactly the same kind of $\frac{Q}{p\cdot q} j$
    factor for each
    internal leg of each vertex that we had for the external legs.
    Furthermore, in the null zone the internal ratio $\frac{Q}{p\cdot q}$
    will equal the common external ratio by the same rotten fruit
    calculation
    described earlier. These identities have reduced the problem to a sum
    of vertex-source problems, and by the earlier arguments, the null zone
    applies to all trees.

    \subsection*{Gauge invariant reorganization of Feynman diagrams}

    The result of applying the decomposition identities and the contracted
    forms of the external leg attachments is a new reduced and rearranged
    amplitude, a sum over vertices
    \begin{equation}
    M_{\gamma}(T_G) = \sum M_{\gamma}(V_G) R(V_G)
    \end{equation}
    with the vertex attachments separated out from the rest of the graph
factors
    $R$.
    Each term in the sum is separately gauge invariant.
    The radiation zero is evident from the fact that
    each factor $M_{\gamma}(V)$ vanishes in the null zone.

    \paragraph*{Alternate theorem.}

    Notice that charge conservation $\sum Q_i = 0$
    is dual to Eq.(6) and implies
    \begin{equation}
    M_{\gamma}(T_G) = 0 \;\;\; {\rm if\;} \frac{J_i}{p_i\cdot q} = {\rm same}
    \end{equation}
    This alternate interference theorem refers to zeros that, by contrast,
    are spin-dependent but independent of charge.

    \subsection*{Radiation symmetry and representation}
    Both theorems can be stated as symmetries.\cite{B}  They correspond to
    invariance under either replacement,
    \begin{equation}
    \frac{Q_i}{p_i\cdot q} \rightarrow \frac{Q_i}{p_i\cdot q} + C
    \;\;{\rm or}\;\;
    \frac{J_i}{p_i\cdot q} \rightarrow \frac{J_i}{p_i\cdot q} + C'
    \end{equation}
    By the appropriate choice of $C$ and $C'$, all zeros can be made
    explicit in a ``radiation representation'' for the vertex amplitudes:
    \begin{equation}
    M_{\gamma}(V_G) = \sum p_i\cdot q (\frac{Q_i}{p_i\cdot q}-
    \frac{Q_j}{p_j\cdot q})(\frac{J_i}{p_i\cdot q} -
    \frac{J_k}{p_k\cdot q})
    \end{equation}
    for $i \ne j,k$.
    This reduces $M_{\gamma}(T_G)$
    to a single factored form for the simplest case, $n=3$, in agreement
    with the original work by Goebel, Halzen and Leveille.\cite{GHL,DZ}
    Their beguiling statement for the original
    $q\overline{q} \rightarrow W^{\pm}\gamma $ is that the three Feynman
    amplitudes reduce to the sum of two Abelian (QED-like)
    diagrams multiplied by a factor in which the zero is explicit.

    \section*{Multi-Photons and General Decoupling Theorem}

    What about neutrals amongst the $n$ particles, including other photons?
    The quick answer from an examination of the $Q/p\cdot q$ factor is
    the right one.  All neutral particles must be massless and travel in the
    same direction as the photon.  Strictly, in the proof we must have the
    corresponding current $J_i$ vanish in the null zone, and this is what is
    found.\footnote{For very special cases, like Compton amplitudes, there
    are no radiation zeros, physical or non-physical, if the null zone
    corresponds to forward scattering of massless neutral vector particles.}

    In fact, the
        multiphoton zeros and the connection to local Poincare
    transformations follow from a decoupling theorem~\cite{BK,BK2} for
        the scattering of a system of particles immersed in an external
        electromagnetic plane wave.
    On the way to the theorem it is first shown that for a particle coupled
    to an external electromagnetic plane wave
    the wave functions for spins $0$, $1/2$, $1$
    all can be written in the form
        \begin{equation}
        \Psi (x)=ULT\chi (x)
        \label{eq:psi}\end{equation}
    where $\chi$ is the free solution.
    The $U$,$L$,$T$ are local gauge,
        Lorentz, and displacement transformations, respectively, whose
    first-order terms are exactly the ones we have been talking about.

    The identity used to show that these are indeed solutions ends up
    being a cornerstone to the whole radiation zero business.  It is
     \begin{equation}
    (UT)^{-1}D^{\mu}UT = \Lambda^{\mu \nu}\partial_{\nu}
    \label{eq:identity}\end{equation}
    in which $D$ is the covariant derivative (hiding the plane wave inside
    it) and $\Lambda$ is the finite
    Lorentz transformation (little group element, actually) corresponding to
    ye olde $\omega$.  {\it It is observed that the plane wave has been
    swallowed up into a local space-time symmetry of the equation of
    motion.}  The $\Lambda$ terms go away by invariance, and isn't this
    familiar by now?

    To finish the story of the theorem, consider the tree amplitude for the
        scattering of a system of particles with no external field.
        If we turn on an external
        electromagnetic field, the internal and external legs of the
    tree
        amplitude are altered according to the Fourier
    transforms of the $ULT$ factors.  In the null zone, all factors collapse
    to unity from charge conservation, Lorentz invariance, and momentum
    conservation.  I have not done justice to the whole story, but the
    bottom line is that, like Perseus, the system of particles can be
    invisible to an external plane wave in special regions of phase space.

    \section*{When Bad Things Happen to Good Zeros}

    Like some other recently well-publicized evidence, measurements of
    radiation zeros are destined to be contaminated, compromised and
    ultimately corrupted. See Fig. 2.
    To start with, higher-order corrections
    will not vanish in the null zone.  The internal factors $Q/p\cdot q$ in
    closed loops correspond to momenta $p$ that is integrated over; these
    factors are certainly not fixed by the outside legs.  {\it It is
    interesting, however, that radiation symmetry Eqs.(10) still hold for the
    complete quantum amplitude to all orders in perturbation theory.}

    Anomalous photon couplings spoil the zeros in lowest order.
    Many speakers at this conference have focused on limits that can
    be set on the $\kappa$ and $\lambda$ parameters for both
    $WW\gamma$ and $WWZ$ trilinear couplings.  As indicated by the earlier
    remarks, the $W\gamma$ zero provides
    litmus tests for these parameters.  Any deviations from the gauge theory
    values lead to momentum dependence in the photon attachment currents
    such that the first-order transformations can no longer be universal.

    Higher derivatives than the ones announced as ``gauge-theoretic''
    produce terms in the $J$ currents that are higher order in $q$.  These
    terms have {\it a priori} no mechanism, no additional symmetry to effect
    their cancellation in a sum over all the legs of a given vertex.  The
    Yang-Mills
    $\cal{O}$$(q^2)$ is the one exception that proves the rule.

    \begin{figure} 
       \caption{But most zeros are not in the physical phase space.}
        \end{figure}

    \subsection*{Generalization to a Larger Gauge Sector}

    There are zeros associated with any gauge group when the corresponding
    massless gauge bosons are emitted.  The ``charges'' now are the
    Clebsch-Gordan coefficients coming from
    the attachment of a boson belonging to the adjoint representation of the
    gauge group. The bad things here are two-fold.  In QCD, color charges
    are averaged or summed over in hadronic reactions.  The zeros are for
    the most part washed out, even when perturbation theory is applicable
    (in deep inelastic reactions, and so forth).  In thinking of the weak
    bosons themselves, electroweak symmetry is broken.  {\it Radiation zeros
    require the internal symmetry to remain good:  The charges `` $Q_i$''
    must be conserved.}  Of course, the weak-boson masses are far from
    vanishing; a nonvanishing $q^2$ itself ruins radiation interference.

    It is not surprising that in a supersymmetric or extended supersymmetric
    world, the photonic zeros would have partner photino zeros and
    ``sphotino'' zeros.  There are ``szeros'' and ``xeros'' in the
    supersymmetrically extended gauge sector.\cite{BK3,DGT}  The well-known
    problem is that SUSY is experimentally evasive and
    hardly unbroken.

    \section*{Recent Work}

    An ``approximate zero'' in $WZ$ production by fermion-antifermion
    annihilation has been proposed recently by Baur, Han and
    Ohnemus\cite{BHO,TH} as another test of the self-couplings.
    At high energies,  the sensitivity to the $WWZ$ vertex is not so
    terribly different than that of the
    $W\gamma$ channel on the $WW\gamma$ vertex.
    It was briefly noted in the early, detailed paper on radiation
zeros\cite{BKB}
    that the dip structure found\cite{BSM} in the $WZ$ angular distributions
    corresponds to an approximate radiation interference zero.
    Baur {\it et al.} show that even in the high-energy limit there is a
    nonvanishing longitudinal helicity amplitude.  Even if the c.m. energy
    is large compared with the masses, the couplings still refer to a broken
    symmetry theory and the zero will remain approximate at high energy.
    The approximate zero corresponds to broken radiation symmetry,
    nevertheless, and the origin and mechanism for the partial cancelations are
    the same as for the exact zeros.

    Some day, experiments on multiboson zeros, exact or approximate, will make
    sense.  Recent discussions of multiphoton/gluon reactions\cite{DIA,BCS}
    could be extended to the broken symmetry weak-boson production channels.
    In $WZZ$ production, for example, quadrilinear couplings come into play,
    along with a multi-$Z$ approximate zero.  But whether these couplings
    can be probed, even in the next generation of colliders, is very much
    open to question.

    \section*{Right and Left}

    \subsection*{What is right}
    It may seem reasonable to say that the radiation zero phenomena are
    well understood.   Among other things, they are another property of
    gauge theories.  We have the following nice connections.

    \paragraph*{Classical interference.}

    The zeros are the generalization of the well-known
    vanishing of classical nonrelativistic electric and magnetic dipole
    radiation occurring for equal charge/mass ratios (indeed, the low-energy
    limit of the null zone conditions!) and gyromagnetic
    g-factors.  The
    null zone is exactly the same as that for
    the completely destructive interference of
    radiation by charge lines (a classical
    convection current calculation\cite{BKB}) and is preserved
    by the fully relativistic quantum Born approximation for gauge theories.

    \paragraph*{Gauge couplings as transformations.}

    Proving radiation symmetry theorems has brought forth the fact that
    gauge boson couplings to particles, including self-couplings,
    can be interpreted as transformations of the
    particles in both internal and external space.
    This is understandable since the gauge bosons belong to adjoint
    representations of both the Lie gauge groups and the Poincare space-time
    group.  Only for the gauge couplings, however, do we get the universal
    Poincare generator representation by the spin and contact currents,
    which are necessary for radiation symmetry.  For the SUSY and extended
    SUSY cases, we generate universal local supersymmetry and chiral
    transformations, respectively.

    \paragraph*{The long and the short of it.}

    We have noted that the gauge-theory Born amplitudes have the
    same null zone as the classical radiation patterns. In the
    short-distance limit, gauge theories can be renormalized, corresponding
    to good high-energy behavior for the Born amplitudes.  Thus, only for
    couplings that correspond to $g=2$, for example, do both the short and
    long distance behaviors fall into these special categories.  (Brodsky
    and Schmidt\cite{BS} emphasize the magic of $g=2$, including its
    implications for photoabsorption sum rules.)
    We have this connection between the small and large distance scales.

    \subsection*{What is left}

    Still, there are to me, at least, some loose ends.

    \paragraph*{Spin barrier.}

    It has not been possible to find a theory of photon couplings, for
    instance, to
    particles with spins greater than one that preserves radiation symmetry.
    I at least have failed thus far in attempts to use
    supermultiplets in the gauge and matter sectors motivated by
    supersymmetry and string theories.  Passarino tells how gravitons spoil
    radiation symmetry, and how their radiation does not seem to have any
    analogous zeros.\cite{P}  One way of describing this
    theoretical wall is in terms of a power series in photon momentum.
    (modulo the $Q/p\cdot q$ factor).  The zeroth and first order terms are
    controlled by translation and Lorentz symmetry, respectively.  The
    isolated instance of second-order terms in the Yang-Mills source vertex
    is controlled by the Bianchi identity, reminiscent of a
    curved-space-time symmetry.  Higher spins lead to additional
    second-order and higher-order terms for which there must be new
    symmetries controlling them.

    \paragraph*{Mixing internal and external spaces.}

    The null zone is defined by equations mixing internal charges and
    phase space.  The original $W\gamma$ zero occurs at angles given in
    terms of quark charges.  Radiation
    symmetry can be rewritten
    $Q_i \rightarrow Q_i + C p_i \cdot q$.
    The mixing together of internal and external parameters suggests a look at
ideas
    such as
    those behind Kalusza-Klein theories where a fifth or higher dimension is
    defined in terms of charges and gauge fields.  The radiation symmetry
    could be part of the larger space-time symmetry, and radiation zeros
    a decoupling of the extra coordinate(s).

    \section*{Acknowledgment}
    Many thanks go to Ulrich Baur, Tao Han and Thomas Muller for their help
    and hospitality.  I am grateful to Ken Kowalski for discussions on
    these matters through the years.


\begin{references}
\bibitem{EXP} The progress in the experimental study of trilinear
    couplings through the production of all electroweak pairs at Fermilab
    has been the subject of talks by Aihara, Fuess, Johari, Landsberg,
    Neuberger, Wagner, and Zhang at this conference.  See those papers for
    references.
\end{references}
        \end{document}